\documentclass[conference,a4paper]{IEEEtran}
\IEEEoverridecommandlockouts

\usepackage[hidelinks]{hyperref}
\usepackage[cmex10]{amsmath}
\usepackage{amssymb,amsfonts}
\usepackage{dblfloatfix}

\usepackage[ruled,vlined]{algorithm2e}
\usepackage{graphicx}
\graphicspath{{Figures/PDF/}{Figures/PNG/}}

\usepackage{booktabs}
\usepackage{siunitx}
\usepackage[numbers,compress]{natbib}
\usepackage{texnames}
\usepackage{bm,bbm}
\usepackage{orcidlink}

\begin{document}

\title{\uppercase{Coverage Path Planning For Multi-view SAR-UAV Observation System Under Energy Constraint}
\thanks{*Corresponding author: Xiangyin Zhang.}
}

\author{\IEEEauthorblockN{Deyu Song\orcidlink{0009-0007-5931-9365}}
		\IEEEauthorblockA{\textit{University of Electronic Science and Technology of China}\\
		611731 Chengdu, China\\
		dysong@std.uestc.edu.cn}\\
		
		\IEEEauthorblockN{Zipei Yu}
		\IEEEauthorblockA{\textit{University of Electronic Science and Technology of China}\\
		611731 Chengdu, China\\
		202221100103@std.uestc.edu.cn}\\
	\and
		\IEEEauthorblockN{Xiangyin Zhang*\orcidlink{0000-0003-0435-9582}}
		\IEEEauthorblockA{\textit{University of Electronic Science and Technology of China}\\
		611731 Chengdu, China\\
		zhangzxy@uestc.edu.cn}\\
		
		\IEEEauthorblockN{Kaiyu Qin\orcidlink{0000-0002-0812-7095}}
		\IEEEauthorblockA{\textit{University of Electronic Science and Technology of China}\\
		611731 Chengdu, China\\
		kyqin@uestc.edu.cn}\\
}

\maketitle
\begin{abstract}
	Multi-view Synthetic Aperture Radar (SAR) imaging can effectively enhance the performance of tasks such as automatic target recognition and image information fusion. Unmanned aerial vehicles (UAVs) have the advantages of flexible deployment and cost reduction. A swarm of UAVs equipped with synthetic aperture radar imaging equipment is well suited to meet the functional requirements of multi-view synthetic aperture radar imaging missions. However, to provide optimal paths for SAR-UAVs from the base station to cover target viewpoints in the mission area is of NP-hard computational complexity. In this work, the coverage path planning problem for multi-view SAR-UAV observation systems is studied. First, the coordinate of observation viewpoints is calculated based on the location of targets and base station under a brief geometric model. Then, the exact problem formulation is modeled in order to fully describe the solution space and search for optimal paths that provide maximum coverage rate for SAR-UAVs. Finally, an Adaptive Density Peak Clustering (ADPC) method is proposed to overcome the additional energy consumption due to the viewpoints being far away from the base station. The Particle Swarm Optimization (PSO) algorithm is introduced for optimal path generation. Experimental results demonstrate the effectiveness and computational efficiency of the proposed approach. 
\end{abstract}


\section{Introduction}

Synthetic Aperture Radar (SAR) imaging has the advantages of all-day and all-weather operation, strong penetration, high resolution, and has been playing an important role in applications such as military surveillance and civil management \cite{cao_automatic_2016}. Limited to the imaging mechanism, there might be blind areas for a single-view synthetic aperture radar imaging system \cite{9553308}. With the rapid development of computer technology and unmanned aerial vehicles (UAVs) technology, the UAV-borne synthetic aperture radar imaging system has been introduced to overcome the above problem. UAVs are characterized by low cost, light weight, and high flexibility. A swarm of UAVs with synthetic aperture radar imaging equipment is able to observe targets from different views. If appropriately deployed, such multi-view SAR-UAV observation system can not only fix the existence of blind areas, but capable of obtaining higher resolution images with less noise through data fusion and resolution enhancement operations \cite{7515203}.

The multi-view SAR-UAV observation system has been proved to be effective in enhancing the performance of tasks such as automatic target recognition \cite{9520257} and image information fusion \cite{8340833}. However, UAVs have a limited amount of energy, and how to utilize that limited energy to efficiently perform observation missions is pivotal. In multi-view SAR-UAV observation missions, UAVs with limited energy are required to cover several targets from different viewpoints, which can be described as a Coverage Path Planning (CPP) problem. The goal of the CPP problem is to find optimal paths for UAVs that cover as many viewpoints as possible under an energy constraint. Different kinds of constraints and various mission requirements need to be considered in the CPP problem, which makes it a multi-objective and multi-constraint optimization problem. In terms of optimization, the CPP problem is an NP-hard complexity one \cite{869506}.

Meta-heuristic based approaches are usually inspired by natural group intelligence, physical mechanisms, or evolution theory, and are able to search globally to find an approximate optimal solution for CPP problem without a time-consuming training process. Meta-heuristic based methods such as Genetic algorithm (GA) \cite{9098989}, Particle Swarm Optimization (PSO) \cite{9402352}, and Ant Colony Optimization (ACO) \cite{app9071470} have been shown to efficiently solve the CPP problem. Nevertheless, most of the existing approaches on the CPP problem for multi-view SAR-UAVs are designed for a single target \cite{9884308} or an independent polygonal area \cite{9987002}. There is a lack of research on the swarm-scale multi-view SAR-UAV observation system.

The clustering mechanism is a flexible and efficient method for simplifying the CPP problem by grouping target viewpoints into clusters. The clusters can then be allocated to UAVs so that the computational complexity of the CPP problem will be greatly reduced. In this work, an Adaptive Density Peak Clustering (ADPC) method is proposed to group target viewpoints, and approximate optimal paths that maximize the coverage rate for SAR-UAVs are obtained through meta-heuristic based algorithm. The proposed coverage path planning method can achieve a higher coverage rate compared to existing methods under energy constraint.

\section{Methodology}

In this section, the coordinates of viewpoints are generated based on a brief geometric model, and the exact problem formulation of the multi-view SAR-UAV CPP problem is described. The proposed ADPC method is utilized to solve the CPP problem.

\subsection{Viewpoint Generation}

Consider $M$ targets located in a specific mission area required to be observed by $N$ SAR-UAVs in $K$ different observation views. We assume that all SAR-UAVs have the same flight altitude $H$ and imaging pitch angle $\theta$. The coordinate of the targets can be denoted as $L=\left[ \left({{x}_{1}},{{y}_{1}},0\right),\ldots,\left({{x}_{m}},{{y}_{m},0} \right),\ldots,\left({{x}_{M}},{{y}_{M},0}\right)\right]$, and the coordinate of the base station is set as $\left(0,0,0\right)$. The coordinate of viewpoints for the $m{-}th$ target can be calculated as follows.
\begin{figure}[t]
	\centering
	\includegraphics[width=5cm]{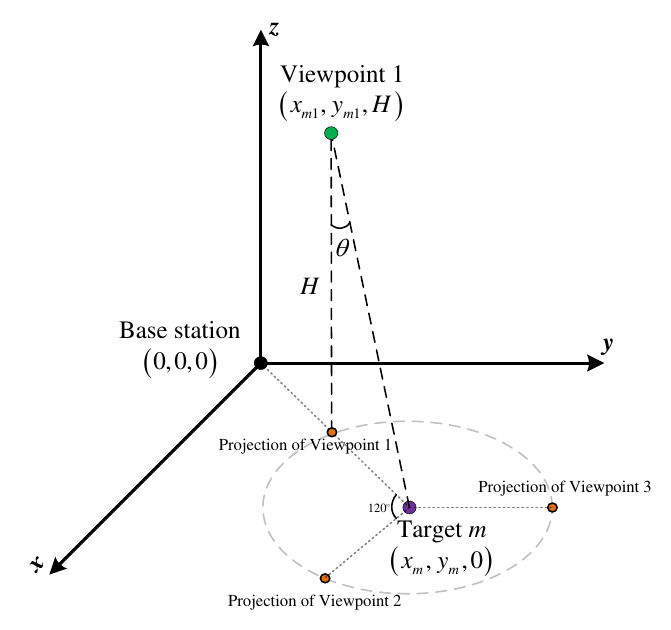}
	\caption{Geometric model of viewpoint generation $\left(K=3\right)$.}\label{fig:model1}
\end{figure}

In this work, the first viewpoint should be located nearest to the base station while satisfying the flight altitude and imaging pitch angle, as shown in Fig.~\ref{fig:model1}. And the horizontal coordinate of the first viewpoint $\left({{x}_{m_1},{y}_{m_1}}\right)$ can be expressed as:
\begin{equation}
\begin{matrix}
		{{x}_{m_1}}=\left( 1-\frac{H\cos \theta }{\sqrt{x_{m}^{2}+y_{m}^{2}}} \right) \cdot {{x}_{m}}  \\
		{{y}_{m_1}}=\left( 1-\frac{H\cos \theta }{\sqrt{x_{m}^{2}+y_{m}^{2}}} \right) \cdot {{y}_{m}}  \\
\end{matrix}
\end{equation}

In this work, we hope that the angle between different viewpoints is as large as possible. Therefore, taking the target $m$ as the center of the circle, the angle between the viewpoint projection points should be $\alpha=360/K$. And the horizontal coordinate of the $k{-}th$ viewpoint of the $m{-}th$ target $(x_{m_k},y_{m_k})$ can be calculated as:
\begin{equation}
\begin{small}
{\begin{matrix}
		{{x_{m_k}} = ({x_{m_1}} - {x_m}) \cdot \cos ({\alpha _k}) - ({y_{m_1}} - {y_m}) \cdot \sin ({\alpha _k}) + {x_m}}\\
		{{y_{m_k}} = ({x_{m_1}} - {x_m}) \cdot \sin ({\alpha _k}) + ({y_{m_1}} - {y_m}) \cdot \cos ({\alpha _k}) + {y_m}}
\end{matrix}}
\end{small}
\end{equation}
where ${\alpha _k} = \alpha \cdot (k-1),k \in (1,K],k \in {\rm Z}$. The coordinates  of all viewpoints in the mission area can hereby be found by calculation as above.

\subsection{Exact Problem Formulation} 

In this subsection, we present an exact formulation of the CPP problem for multi-view SAR-UAV system. In this work, the total objective is defined as maximizing the overall coverage rate subject to multiple constraints. UAVs should start from the base station, observe as many viewpoints as possible within limited energy capacity, and finally return to the base station. When covering observation viewpoints, the flight paths of UAVs should satisfy the following constraints.
\begin{itemize}
	\item Energy capacity constraint. Due to limitations of load capacity, battery technology, and imaging ability, the energy consumption of SAR-UAVs needs to be considered. The total energy consumption on flight distance and synthetic aperture radar imaging should not be larger than its maximum energy capacity, which can be expressed as:
	\begin{equation}
		{E_F} \cdot {d_i} + {E_I} \cdot {M_{i{-}covered}} \le {E_{\max }}
		\label{constraint1}
	\end{equation}
	where $i \in [1,N],i \in {\rm Z}$. ${E_F}$ and ${E_I}$ denote the energy consumption per unit flight distance and energy consumption to cover each viewpoint, respectively. ${d_i}$ represents the total flight distance of the $i{-}th$ UAV, and $M_{i{-}covered}$ is the number of corresponding covered viewpoints. $E_{\max}$ is the maximum energy capacity.
	\item Detection requirement constraint. In this work, we assume each viewpoint should be detected by only one UAV in order to ensure task implementation quality and avoid resource squandering. Assume that $P=\{ P_1,P_2,\ldots,P_N \}$ is the generated flight paths for $N$ UAVs, there has
	\begin{equation}
		\forall m_k \in {P_i},m_k \notin {P_j}
		\label{constraint2}
	\end{equation}
	where $m \in [1,M],k \in [1,K],i,j \in [1,N],m,i,j,k \in {\rm Z}$.
	\item Overall task requirement constraint. It must be ensured that at least one viewpoint has been covered for each target. This constraint ensures the validity of the overall mission and avoids having targets unobserved, which can be expressed via,
	\begin{equation}
		\forall m \in [1,M],\exists k \in [1,K], \exists i \in [1,N],{m_k} \in {P_i}
		\label{constraint3}
	\end{equation}
	where $m,k,i \in {\rm Z}$.
\end{itemize}

Therefore, the multi-view SAR-UAV coverage path planning problem can be stated as:
\begin{equation}
\begin{array}{l}
	\max \quad \frac{{{M_{{{covered}}}}}}{{M \cdot K}}\\
	s.t.\quad (\ref{constraint1}),(\ref{constraint2}),(\ref{constraint3})
\end{array}
\end{equation}
where $M_{{{covered}}}$ denotes the total number of covered viewpoints. To simplify the problem formulation for ease of solution, the physical constraints of UAV flight are not considered in this work.

\subsection{Adaptive Density Peak Clustering}

In order to solve the above formulation more effectively, we improve the Density Peak Clustering (DPC) \cite{1242072} algorithm based on our scene representation. The specific process of the proposed method is as follows.

\subsubsection{Cluster Center Selection}

The selection of cluster centers in the proposed ADPC method depends on two factors, viewpoint relative density $\rho$ and distance factor $\delta$. In this work, the relative density $\rho_{m_k}$ of viewpoint $m_k$  can be calculated through Gaussian core, which can be expressed as:
\begin{equation}
	{\rho _{{m_k}}} = \sum\limits_{m \ne n} {\exp \left[ { - {{\left( {\frac{{D({m_k},{n_l})}}{{{d_c}}}} \right)}^2}} \right]}
\end{equation}
where $D({m_k},{n_l})$ denotes the distance between viewpoint ${m_k}$ and ${n_l}$. $d_c$ is the cutoff distance, which can be empirically derived. 

The distance factor $\delta_{m_k}$ of viewpoint $m_k$ can be achieved by calculating its minimum distance from viewpoints of higher density. In the case of the viewpoint with the highest relative density, its distance factor takes its maximum distance from any other viewpoints. Therefore, the distance factor $\delta_{m_k}$ can be calculated as:
\begin{equation}
	\begin{small}
	   {\delta _{{m_k}}} = \left\{ {\begin{array}{*{20}{c}}
		{\min [D({m_k},{n_l})]{\kern 1pt} }&{if\;{\rho _{{m_k}}} \ne {\rho _{\max }},{\rho _{{n_l}}} > {\rho _{{m_k}}}}\\
		{\max [D({m_k},{n_l})]{\kern 1pt} }&{if\;{\rho _{{m_k}}} = {\rho _{\max }}\;\;\;\;\;\;\;\;\;\;\;\;\;\;\;\;\;\;}
		\end{array}} \right.
	\end{small}
\end{equation}

In this work, the cluster centers are selected by choosing $N$ viewpoints with maximum $\gamma=\rho \cdot \delta$, and the cluster center selection phase is completed.

\subsubsection{Clustering Strategy}

After the cluster centers are selected, the rest of viewpoints should be classified into clusters. In this work, an adaptive factor $\sigma$ is introduced to help viewpoints find suitable clusters. 

Obviously, the cluster with a greater distance from the base station will result in higher flight energy consumption for the UAVs. Therefore, when a viewpoint chooses its cluster, not only its distance from the cluster centers should be considered, the distance between the cluster center and the base station, as well as the number of viewpoints in the current cluster, should be taken into account in the clustering strategy. The clustering process starts from viewpoints with a greater distance from the base station. Use $C_i$ to represent the $i{-}th$ cluster center, and the adaptive factor $\sigma$ of viewpoint $m_k$ on center $C_i$ can be expressed as:
\begin{equation}
	{\sigma _{{m_k},{C_i}}} = D({m_k},{C_i}) \cdot {\omega _i}^{\xi  \cdot {r_i}}
\end{equation}
where $D({m_k},{C_i})$ is the distance between viewpoint $m_k$ and center $C_i$. ${\omega _i}$ is the current number of viewpoints in $C_i$. $\xi$ is the expansion index, which is used to balance the parameter range. ${r_i}$ is the relative distance ratio which can be calculated via,
\begin{equation}
	{r_i=\frac{{D({C_i},B)}}{{\frac{1}{N} \cdot \sum\limits_{i = 1}^N {D({C_i},B)} }}}
\end{equation}
where ${D({C_i},B)}$ represents the distance between center $C_i$ and the base station.

Finally, viewpoints should choose the cluster center with the minimum adaptive factor $\sigma$ as their final classified cluster, and the clustering phase can be completed. With our proposed clustering strategy, clusters with a greater distance from the base station tend to contain fewer viewpoints, which can effectively offset the energy consumption disadvantage of clusters being far from the base station.

\subsubsection{Optimal Path Generation}

After clusters have been allocated to SAR-UAVs, optimal paths need to be generated through optimization algorithms. In this work, the PSO algorithm is introduced to obtain optimal paths for SAR-UAVs. The flight distance is set to be the fitness function. After a population of random particles (random solutions) is initialized, an optimal solution can then be found by iteration through following the track of the local best solution and the global best solution. The following experimental results demonstrate that PSO is able to achieve feasible solutions with accuracy and robustness while having acceptable computational speed.

\section{Experimental Results}

In this section, the performance of the proposed ADPC coverage path planning method for multi-view SAR-UAV system is thoroughly evaluated. In addition to the PSO algorithm applied in our proposed method, we also applied the GA algorithm and the ACO algorithm on the proposed ADPC clustering method. Furthermore, we applied the PSO algorithm on the original DPC method and the traditional clustering method, Kmeans. The above four methods ADPC-GA, ADPC-ACO, DPC-PSO, and Kmeans-PSO are introduced for comparison. 

\begin{figure*}[hbt]
	\begin{minipage}[b]{.16\linewidth}
		
		\vspace{3pt}
		\centering
		\centering\includegraphics[width=3cm]{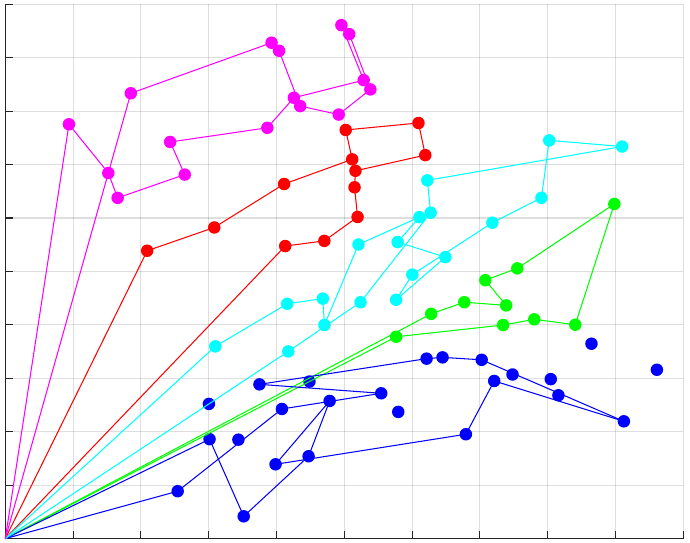}\\
		\centering (a)DPC-PSO.	
	\end{minipage}
	\hfill
	\begin{minipage}[b]{.16\linewidth}
		
		\vspace{3pt}
		\centering
		\centering\includegraphics[width=3cm]{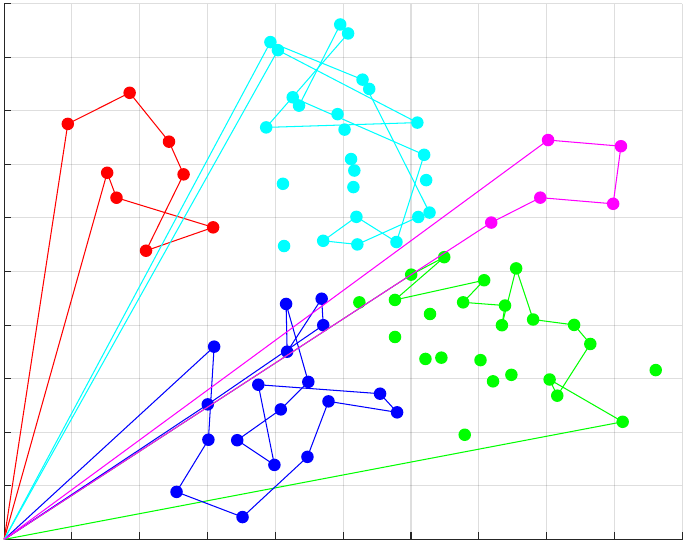}\\	
		\centering (b)Kmeans-PSO.
	\end{minipage}
	\hfill
	\begin{minipage}[b]{.16\linewidth}
		
		\vspace{3pt}
		\centering
		\centering\includegraphics[width=3cm]{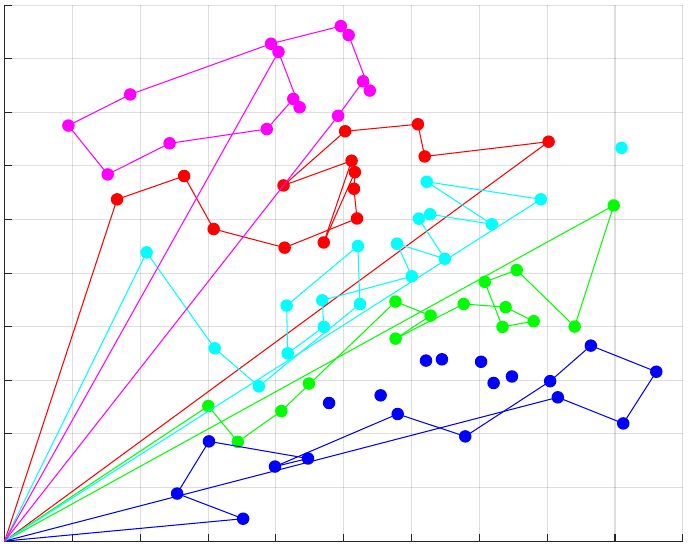}\\	
		\centering (c)ADPC-GA.
	\end{minipage}
	\hfill
	\begin{minipage}[b]{.16\linewidth}
		
		\vspace{3pt}
		\centering
		\centering\includegraphics[width=3cm]{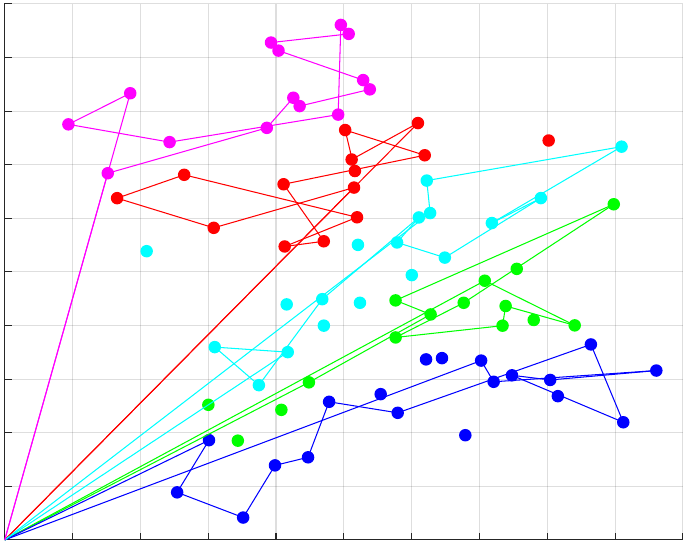}\\	
		\centering (d)ADPC-ACO.
	\end{minipage}
	\hfill
	\begin{minipage}[b]{.16\linewidth}
	
		\vspace{3pt}
		\centering
		\centering\includegraphics[width=3cm]{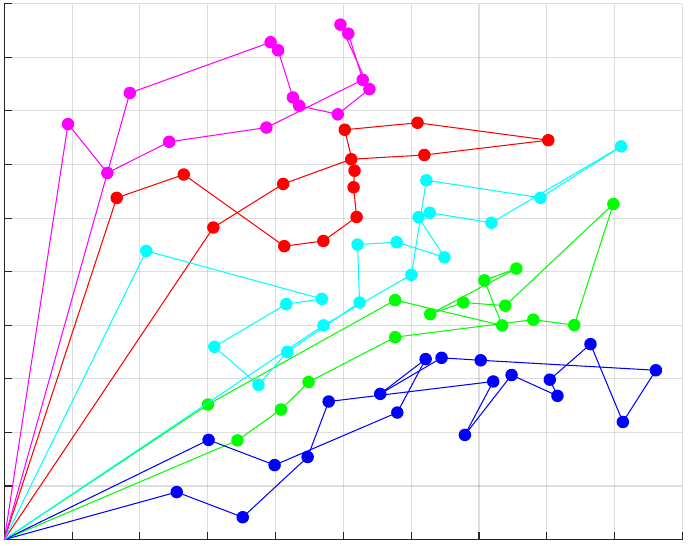}\\	
		\centering (e)Proposed.
	\end{minipage}
	\caption{Optimal paths for SAR-UAVs generated by different CPP methods.}
	\label{fig:views}
\end{figure*}

In our experiment, targets are randomly generated in a mission area of $2000m \times 2000m$ size. The number of targets $M$ is set as 20. There are two variable parameters, the number of UAVs $N$ and the number of viewpoints $K$, and we will experimentally analyze the performance of the coverage path planning methods with one of them fixed and the other changed. Each set of experiments was run for 1000 times, the maximum iterations and population size of PSO, GA, and ACO are set as 100 and 30, respectively.

Fig.~\ref{fig:views} shows visualized optimal flight paths for SAR-UAVs to cover target viewpoints generated by different methods under the same experimental parameter setting. It can be seen that the proposed ADPC-PSO method can achieve paths with full coverage, while the other methods appear to have uncovered viewpoints due to limited energy capacity. This proves the effectiveness of the design of the adaptive factor in the clustering phase. The introduced adaptive factor can well balance the number of viewpoints in the clusters in order to achieve energy equalization for SAR-UAVs.

\begin{figure}[h]
	\centering
	\includegraphics[width=6.3cm]{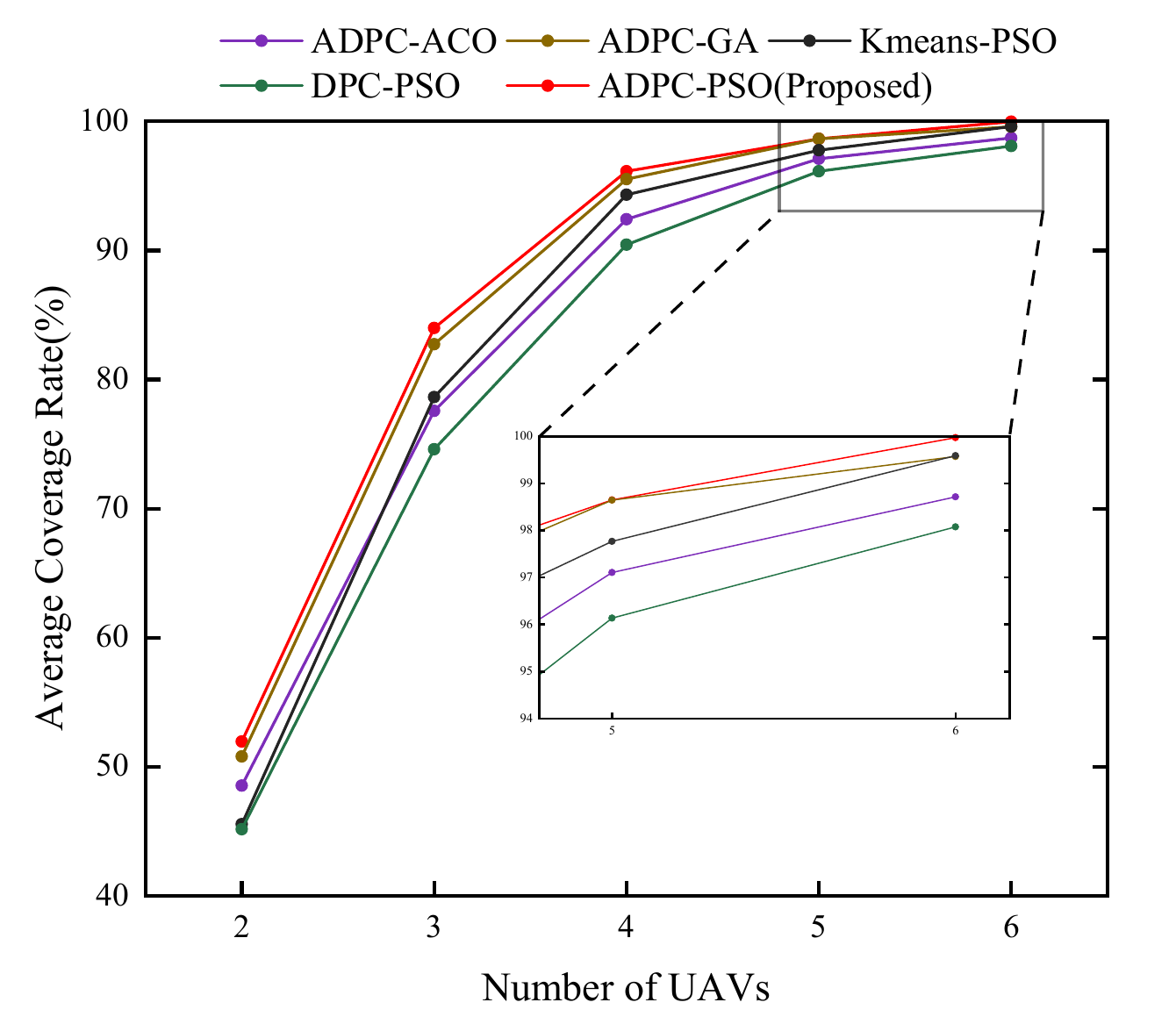}
	\caption{Average coverage rate obtained by CPP methods when different numbers of SAR-UAVs are used.}\label{fig:cover1}
\end{figure}

\begin{figure}[t]
	\centering
	\includegraphics[width=6.3cm]{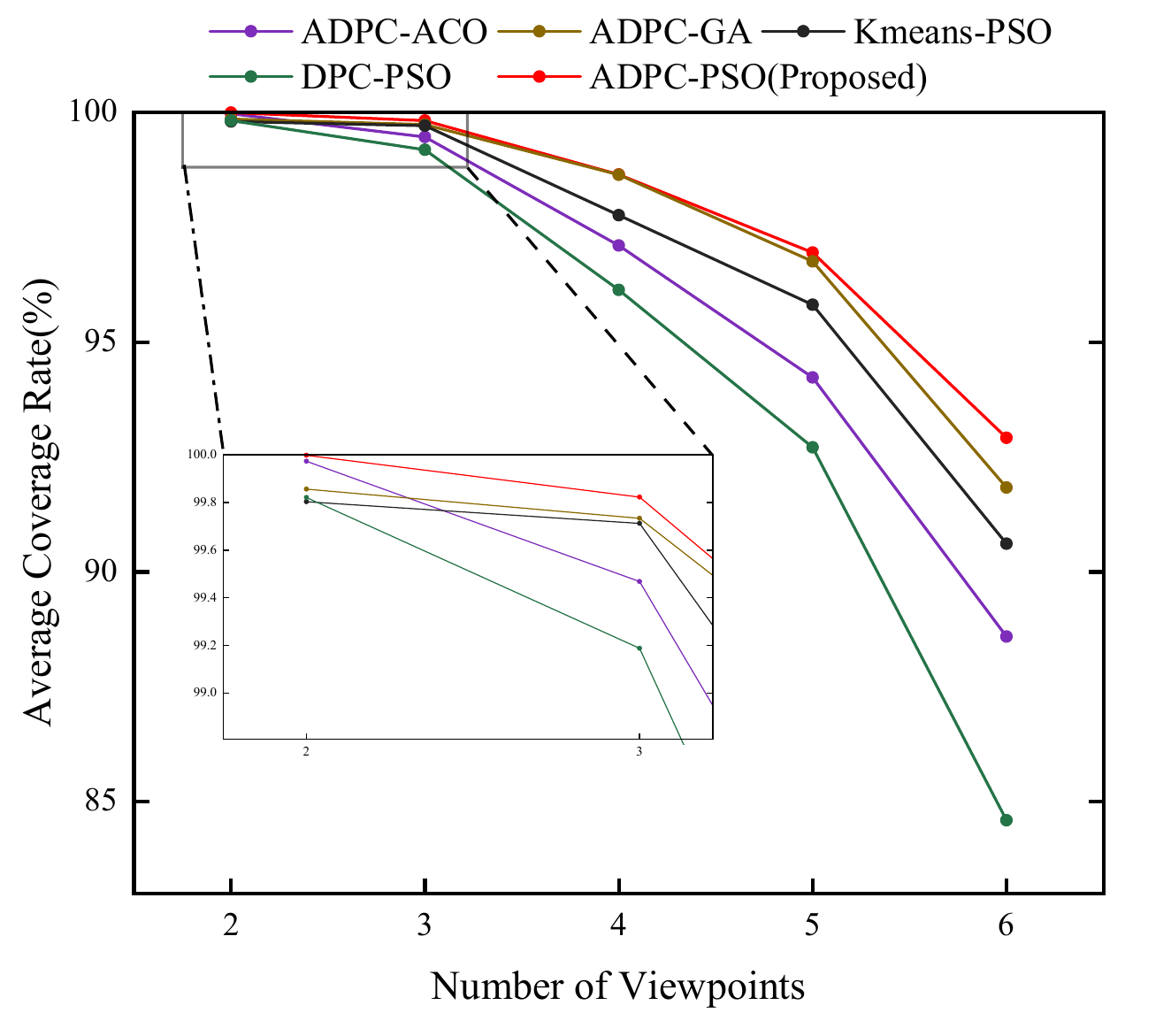}
	\caption{Average coverage rate obtained by CPP methods when different numbers of observation viewpoints are used.}\label{fig:cover2}
\end{figure}

Fig.~\ref{fig:cover1} illustrates the average coverage rate obtained by CPP methods when the number of SAR-UAVs $N = \{2,3,4,5,6\}$. For the same number of targets and observation views, the average coverage rate goes up along the increase of the number of SAR-UAVs. We can find out that our proposed ADPC-PSO method consistently outperforms other CPP methods in terms of average coverage rate. In Fig.~\ref{fig:cover2}, the effect of different numbers of observation viewpoints on the average coverage rate when $N = 5$ is tested. An increase in the number of viewpoints creates more task loads for SAR-UAVs, and the energy capacity can no longer support too many observation tasks. Nevertheless, the proposed method has lower performance degradation in average coverage rate compared to other CPP methods under gradually higher observation task loads. The experiments in Fig.~\ref{fig:cover1} and Fig.~\ref{fig:cover2} demonstrate the effectiveness and superior coverage performance of the proposed coverage path planning method.

The CPU runtime for different clustering strategies and meta-heuristic optimal path generation methods when $M=20$, $N=5$ is as shown in Table~\ref{tab:MyTable}. In terms of CPU runtime, the clustering strategy and optimal path generation method applied in our proposed method are not as good as DPC and GA, respectively. On the other hand, they are still in the same order of magnitude compared to DPC and GA, while outperforming the Kmeans clustering method and ACO algorithm. This proves that the proposed method does not bring excessive computational load and has practical applicability.

\begin{table}[hbt]
	\centering
	\caption{CPU runtime (s) for different clustering strategies and meta-heuristic optimal path generation methods.}\label{tab:MyTable}
	\begin{tabular}{c c c c c c c}
		\toprule
		 & {Kmeans} & {DPC} &  \textbf{ADPC}  & {GA} & {ACO} &  \textbf{PSO} \\ 
		\cmidrule(r){1-1} \cmidrule(lr){2-4}  \cmidrule(l){5-7}
		$K=2$ &	3.043 &	0.284 & 0.702 & 15.988 & 284.016 & 55.026 \\
		$K=3$ & 3.241 &	0.410 & 1.033 & 18.214 & 497.048 & 78.312 \\
		$K=4$ &	3.287 &	0.540 & 1.251 &	19.810 & 743.200 & 105.684 \\
		$K=5$ & 3.446 &	0.705 & 1.585 &	21.126 & 946.052 & 132.578 \\
		$K=6$ &	3.597 &	0.888 & 2.009 & 23.035 & 1155.342 & 162.525\\
		\bottomrule
	\end{tabular}
\end{table}

\section{Conclusion}

This paper focused on the coverage path planning problem of multi-view SAR-UAV system with the aim of finding paths that cover more viewpoints for SAR-UAVs under energy constraint. First, the coordinate of observation viewpoints is calculated based on the location of the targets and the base station. Then, the exact formulation of the coverage path planning problem is established to find optimal paths that provide maximum coverage rate for SAR-UAVs. Finally, an ADPC method is presented to adaptively classify viewpoints into clusters by taking the current number of viewpoints in clusters as well as the distance between the cluster centers and the base station into account. The meta-heuristic algorithm PSO is applied for generating the final optimal paths. Experimental results demonstrate that the proposed method can obtain valid paths for SAR-UAVs that maximize the overall coverage rate within acceptable computation time.

\small
\bibliographystyle{IEEEtranN}
\bibliography{references}

\end{document}